\documentclass[useAMS]{mn2e}
\usepackage{psfig}

\title[BAL Quasars in CFHTLS]{X-ray and optical properties of Broad Absorption Line Quasars in the 
Canada$-$France$-$Hawaii Telescope Legacy Survey}
\author[C. S. Stalin, R. Srianand \& P. Petitjean]{C. S. Stalin$^{1}$\thanks{E-mail:
stalin@iiap.res.in}, R. Srianand$^{2}$ and P. Petitjean$^{3}$ \\ 
$^{1}$Indian Institute of Astrophysics, Block II, Koramangala, Bangalore 560034, India\\
$^{2}$Inter-University Centre for Astronomy and Astrophysics, Post Bag 4, Ganeshkhind, 411007 Pune, India \\
$^{3}$  Institut d'Astrophysique de Paris, CNRS ­ Universit\'{e} Pierre et 
Marie Curie, 98bis bd Arago, Paris 75014, France
}

\begin{document}

\date{Accepted 1988 December 15. Received 1988 December 14; in original form 1988 October 11}

\pagerange{\pageref{firstpage}--\pageref{lastpage}} \pubyear{2008}

\maketitle

\label{firstpage}

\begin{abstract}
We study the X-ray and optical properties of 16 Broad Absorption 
Line (BAL)
quasars detected in a $\approx$3 ~deg$^2$ region common to the wide synoptic (W-1) component 
of the Canada$-$France$-$Hawaii Telescope Legacy Survey (CFHTLS)
and the XMM Large Scale Structure survey (XMM-LSS).
The BAL fraction is found to be 10\% in full sample, 7\% for the optical
colour selected QSOs and as high as
33\% if we consider QSOs selected from their IR colours.
The X-ray detected  non-BAL and BAL quasars have a mean observed
X-ray-to-optical spectral slope ($\alpha_{\rm ox}$) of $-$1.47 $\pm$ 0.13 
and $-$1.66 $\pm$ 0.17 respectively. We also find that the BAL QSOs
have $\alpha_{\rm ox}$ systematically smaller than
what is expected from the relationship between 
optical luminosity and $\alpha_{\rm ox}$ as derived from our sample.
Based on this, we show, as already reported in the literature for quasars
with high optical luminosities, our new sample of BAL QSOs  have X-ray 
luminosity a factor of 
three smaller than what has been found for non-BAL QSOs with similar 
optical luminosities. 
Comparison of hardness ratio of
the BAL and non-BAL QSOs  suggests a possible soft X-ray weakness 
of BAL QSOs. Combining our sample, of relatively fainter QSOs, with others from the 
literature we show that larger balnicity index (BI) and maximum
velocity ($V_{\rm max}$) of the C~{\sc iv} absorption are correlated with
steeper X-ray to optical spectral index.
We argue that this is most likely a consequence of the existence of a lower envelope 
in the distribution of BI (or $V_{\rm max}$) values versus optical luminosity. 
Our results thus show that the previously known X-ray weakness of 
BAL QSOs extends to lower optical luminosities as well.
\end{abstract}

\begin{keywords}
surveys - galaxies: active - quasars: general - Xrays: general
\end{keywords}

\section{Introduction}
Broad Absorption Line (BAL) quasars are Active Galactic Nuclei (AGN)  
characterized by the presence of strong absorption troughs in their
UV spectra.
They constitute an observed fraction of about 10$-$15\% of optically 
selected quasars (Reichard et al. 2003; Hewett \& Foltz 2003). 
Recently, it has been shown that the actual BAL fraction could be
higher as optical colour selection of QSOs may be biased against
the BAL QSOs (see for example, Dai et al. 2008; Shankar et al. 2008; 
Urrutia et al. 2009; Allen et al. 2010)
The BALs are attributed to material flowing outwards from the nucleus with velocities of 
5000 to 50000 km/s (Green et al. 2001). 
These quasars are classified into
three subclasses based on the material producing the BAL troughs. 
High ionization BAL quasars (HiBALs) have broad absorption 
from C~{\sc iv}, Si~{\sc iv}, N~{\sc v} and O~{\sc vi}. 
About 10\% of BAL quasars also show, apart from their HiBAL features, 
broad absorption lines of lower ionization species such 
as Mg~{\sc ii} or Al~{\sc iii} and
are called low-ionization BAL quasars (LoBALs). Finally, LoBALs 
with absorptions from excited states of Fe~{\sc ii} or Fe~{\sc iii} are called
FeLoBALs (Wampler et al. 1995). BAL quasars in general have 
higher optical-UV polarization than 
non-BAL quasars, and the LoBALs tend to have particularly high polarization 
than HiBALs (Hutsemekers et al. 1998; Schmidt \& Hines 1999; 
DiPompeo et al. 2010).  
LoBALs have a more reddened optical continuum compared to non-BAL
and HiBAL QSOs(Becker et al. 2000; Sprayberry \& Foltz 1992), thereby suggesting 
the presence
of larger amounts of dust in them. In 
X-rays too, LoBALs have higher absorbing column densities than HiBALs
(Green et al. 2001; Gallagher et al 2002).

The dichotomy between BAL and non-BAL quasars is often thought to be 
a consequence of orientation. 
The similarity between the optical/UV emission lines
and continuum properties of BAL and non-BAL quasars (Weymann et al. 1991; 
Reichard et al. 2003) 
supports a scenario where the observed fraction of BAL quasars corresponds to the covering
fraction of a wind that could be present in all AGN. 
Earlier spectropolarimetric observations too support this orientation
scheme (Goodrich \& Miller 1995; Hines \& Wills 1995). 
However, recent spectropolarimetric observations
of radio-loud BALs do not favour the orientation dependent scheme
for the BAL phenomenon (DiPompeo et al. 2010). On the other hand, Allen
et al. (2010) found a strong redshift dependence of the  
C~{\sc iv} BAL quasar fraction. They conclude that the
BAL phenomenon cannot be due to an orientation effect only. 
Alternative to the orientation scheme it is argued that the observed BAL
quasar fraction could correspond to the intrinsic fraction of quasars hosting
massive nuclear winds, thereby tracing an evolutionary
phase of the AGN lasting $\approx$10$-$15\% of their lives (Hazard et al. 
1984; Becker et al. 2000; Giustini et al. 2008).
In this scenario, the large amounts of gas and dust surrounding the central
source should lead to enhanced far-infrared and sub-millimeter emission
in BAL quasars with respect to non-BAL quasars (Giustini et al. 2008). However,  
sub-millimeter studies found no (Willott et al. 2003) or 
little (Priddey et al. 2007) differences among the two populations.
Also, the mid-IR properties of BAL and non-BAL quasars of comparable 
luminosities are indistinguishable
(Gallagher et al. 2007). BAL quasars are predominantly radio-quiet
while few radio-loud BAL quasars are also known (Becker et al. 2000; 
Brotherton et al. 2005). 

Although radiative acceleration could be the main driving mechanism 
in BAL quasars (Arav et al. 1994; Srianand et al. 2002), we still do 
not have a clear picture of the physics of outflows/winds
in BALs. X-ray observations can help constrain the physical 
mechanisms  at play in BAL quasar outflows and the different scenarios 
proposed  (Giustini et al. 2008). 
Since the ROSAT survey, BAL quasars which are radio-quiet, have 
been known to have faint soft X-ray to optical luminosity ratio
(Green et al. 1995; Green \& Mathur 1996). Their X-ray luminosity is typically
10$-$30 times lower than expected from their UV luminosity, qualifying
them as soft X-ray weak objects (Laor et al. 1997). This implies that the 
soft X-ray continuum of BAL quasars is either (a) strongly absorbed by
highly ionized material or (b) intrinsically under luminous. 
Given the extreme absorption evident in the ultraviolet, this soft X-ray 
faintness was assumed to result from intrinsic absorption in BAL material 
of high column density, typically $N_{\rm H} > 10^{22}$~cm$^{-2}$ (Gallagher et al. 2006, 
hereafter G06; Green et al. 1995; Green et al. 2001; Brotherton et al. 2005; Fan et 
al. 2009; Gibson et al. 2009). 
However, it has also been argued  that intrinsic X-ray faintness cannot be ruled out as the 
cause for their observed X-ray weakness (Sabra \& Hamann 2001; 
Mathur et al. 2000; Gupta et al. 2003, Giustini et al. 2008; Wang et al. 
2008; Ghosh \& Punsly 2008). 
Radio-loud BAL quasars are also found to be X-ray weak 
when compared with radio-loud non-BAL quasars of similar UV/optical
luminosities (Miller et al. 2009).

X-ray spectral analysis of BAL quasars considering neutral and ionized
absorbers are found to yield low neutral hydrogen 
($N_{\rm H} < 10^{21}$~cm$^{-2}$) and  
high ionized hydrogen ($N_{\rm H}^{i} > 10^{21}$~cm$^{-2}$) column 
density respectively 
(Giustini et al. 2008; Streblyanska et al. 2010). 
It thus seems that the inferred $N_{\rm H}$ values depends on 
(a) the ionization state of the gas in our line of sight to the BAL 
quasar and (b) the absorber either fully or partially covering 
the X-ray source.
Here, we investigate the issue of the X-ray weakness of BAL quasars using 
a new sample of quasars selected in the Canada$-$France$-$Hawaii Telescope
Legacy Survey (CFHTLS\footnote{http://www.cfht.hawaai.edu/Science/CFHTLS/})
and overlapping the XMM Large Scale Structure survey (XMM-LSS) and the Spitzer 
Wide-area InfraRed Extragalactic (SWIRE) Survey  with the aims  of
(i) identifying a homogeneous sample of BAL quasars from a parent quasar sample and 
(ii) of studying the X-ray nature of those identified BAL quasars. 
The sample has been selected without a prior knowledge of the BAL nature of
the objects. This paper is organized as follows. The data set used and the observations 
are described in Sect. 2. Identification of BAL quasars is given 
in Sect. 3. Results of the analysis are presented in Sect. 4 and the conclusions are 
drawn in Sect. 5.  Throughout this paper we adopt a cosmology with 
$H_{\rm o}$~=~70~km~s$^{-1}$~Mpc$^{-1}$, 
$\Omega_{\rm m}$~=~0.27 and $\Omega_{\Lambda}$~=~0.73.

\begin{figure}
\hspace*{-0.5cm}\psfig{file=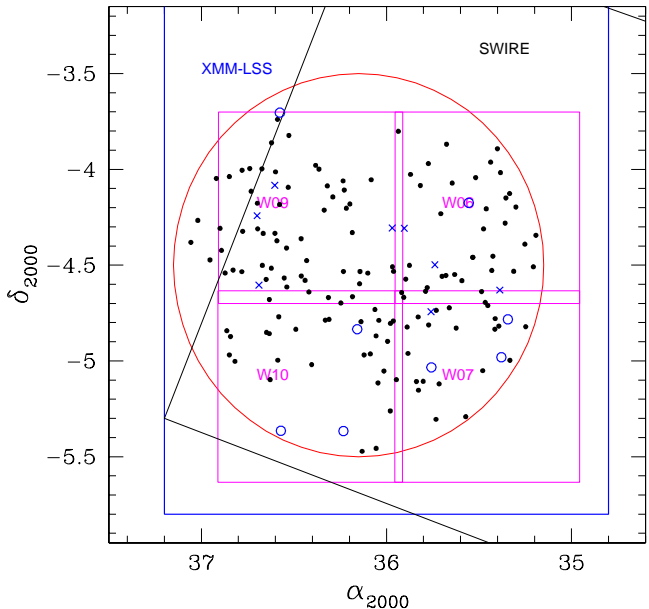,width=9cm,height=9cm}
\caption{Layout of the field used to select quasars in this study.
The XMM-LSS and SWIRE regions are delineated. W06, W07, W09 and 
W10 are the four pointings of CFHTLS and the large circle is the $\sim$3
square degrees region searched for quasars in this work. The quasars
are marked as filled circles, the BAL quasars detected in X-ray are
shown as crosses and the BAL quasars un-detected in X-rays are shown
as open circles.
}
\end{figure}

\section{Dataset}

The optical data set used in this work is  from the wide Synoptic 
component (W-1) of CFHTLS.  CFHTLS images are available in five optical 
bands (u$^{\ast}$, g$^{\prime}$, r$^{\prime}$, i$^{\prime}$ and z$^{\prime}$) 
down to $i_{\rm AB}^{\prime}$~=~24 mag.  The field has been observed in the 
course of the XMM-LSS. Centered at ($\alpha_{2000}$~=~37.5~deg, 
$\delta_{2000}$~=~$-$5~deg), the XMM-LSS (Pierre et al. 2004) is a medium 
depth large area X-ray survey designed to map the large scale structures in 
the nearby universe. The catalog for the first 5.5 square degrees 
(pertaining to 45 XMM pointings) observed in the 0.5$-$2 and 2$-$10 keV  bands,
listing sources above a detection likelihood of 15 in either bands, was 
released by Pierre et al. (2007). Also, overlapping the XMM-LSS and CFHTLS 
fields is the SWIRE Survey (Lonsdale et al. 2003). 

We are in the process of carrying out optical spectroscopic
identification of quasar candidates selected in about 10 square
degrees in the W-1 region. 
Some regions of W-1 also have coverage from XMM-LSS and SWIRE. Therefore
several selection criteria were adopted to select quasar candidates
namely: (i) optical colour-colour criteria similar to that used 
in SDSS (Richards et al. 2002), (ii) IR colour-colour criteria following
Stern et al. (2005) and (iii) optical SED matching using the photometric
redshift code {\it hyperz} (Bolzonella, Miralles \& Pell\`{o} 2000). Our final
list of quasar candidates in W-1 thus consists of (a) candidates selected both by
optical colour and template matching criteria, (b) candidates selected by 
optical colour but missed by template matching method, (c) candidates selected
by optical template matching method but missed by optical colour selection
(d) candidates selected by IR colour selection but missed in optical selection
and (e) all XMM sources having optical counterparts, but not selected
through (a),(b),(c)and (d). Our candidate selection was limited to sources 
brighter than $g^{\prime} <$ 22 mag. 
A complete description
of our procedure of quasar candidate selection as well as the results of
the quasar selection efficiency will be discussed in a forthcoming paper
(Stalin et al. {\it in preparation}) when our complete quasar survey will be reported.
Our main aim here is to understand the X-ray and optical properties of 
BAL quasars. Thus, for this work we have considered only $\approx$3 square
degree region
in W-1, which has both optical and X-ray imaging observations. This
region is shown as a large circle in Fig. 1. Also, marked on this figure
are the regions covered by the XMM-LSS and SWIRE.

Optical spectroscopic observations of the quasar candidates selected
in the $\approx$3 square degrees circular region shown in Fig. 1
were carried out with the AAOmega system (Sharp et al. 2006) on the 3.9 m AAT, 
during two observing runs in September 2006 and 2007. The field of the camera has 
a diameter of 2 degree and was centered at $\alpha_{2000}$~=~36.15 deg and $\delta_{2000}$~=~$-$4.50 deg
(see Fig.~1).  
The 580V and 385R gratings were used, respectively, in the blue and red arms of the spectrograph, 
thereby simultaneously covering the wavelength range 3700$-$8800~\AA, and delivering a spectral 
resolution of $R$$\sim$1300. The total integration time ranged from 30 min to 3 hours depending on the 
brightness of the candidates. 

Spectroscopic data reduction was performed using the AAOmega's data reduction pipeline software  
DRCONTROL. The two dimensional images were flat fielded, and the spectra were extracted (using 
a gaussian profile extraction), wavelength calibrated and combined within DRCONTROL. Redshifts 
were derived using the AUTOZ code (kindly provided to us by Scott Croom). Individual QSO
spectra were manually checked to confirm the correctness of the redshift. Further 
details of the observations and reductions can be found in Stalin et al. (2010).

A total of 159, $z_{\rm em} > 1.5$, new quasars were identified from the 
AAT observations in 
$\approx$3~deg$^2$. They are shown as filled circles in Fig. 1. 
This is the sample we use here to investigate the X-ray properties of BAL
and non-BAL quasars.
Note that the above redshift cut-off is necessary in order to observe the C~{\sc iv}$\lambda$1549 
BAL feature in the spectral range 3800$-$8800~\AA.
Out of these 159 quasars, 120 were primarily selected based on optical colour selection 
without any prior knowledge of X-ray emission, 12 sources were selected based on their IR colour 
(also without any knowledge of X-rays) from the SWIRE survey but were missed by our optical selection
and 27 sources were selected  from the XMM source list through the presence of X-ray emission. 
We find 72\% (i.e. 86 quasars out of 120) of the optically selected quasars are also detected in XMM. 
However, only 33\% (i.e., 4 quasars our of 12) of the IR only selected candidates are detected 
by XMM. All of the 120 optically colour selected quasars were detected by
SWIRE in one of the five IR bands. However, only 78 of these 120 sources
have IR flux values in all the four IR bands so that IR
selection criteria can be applied to them. 
Of these 78, 77 are also IR colour-colour 
selected. The average $u^{\prime}-g^{\prime}$ and $g^{\prime}-r^{\prime}$ 
colours of the 12 IR only selected
quasars are redder than that of the 120 optically colour selected
quasars.

\begin{figure*}
\hspace*{-1.5cm}\psfig{file=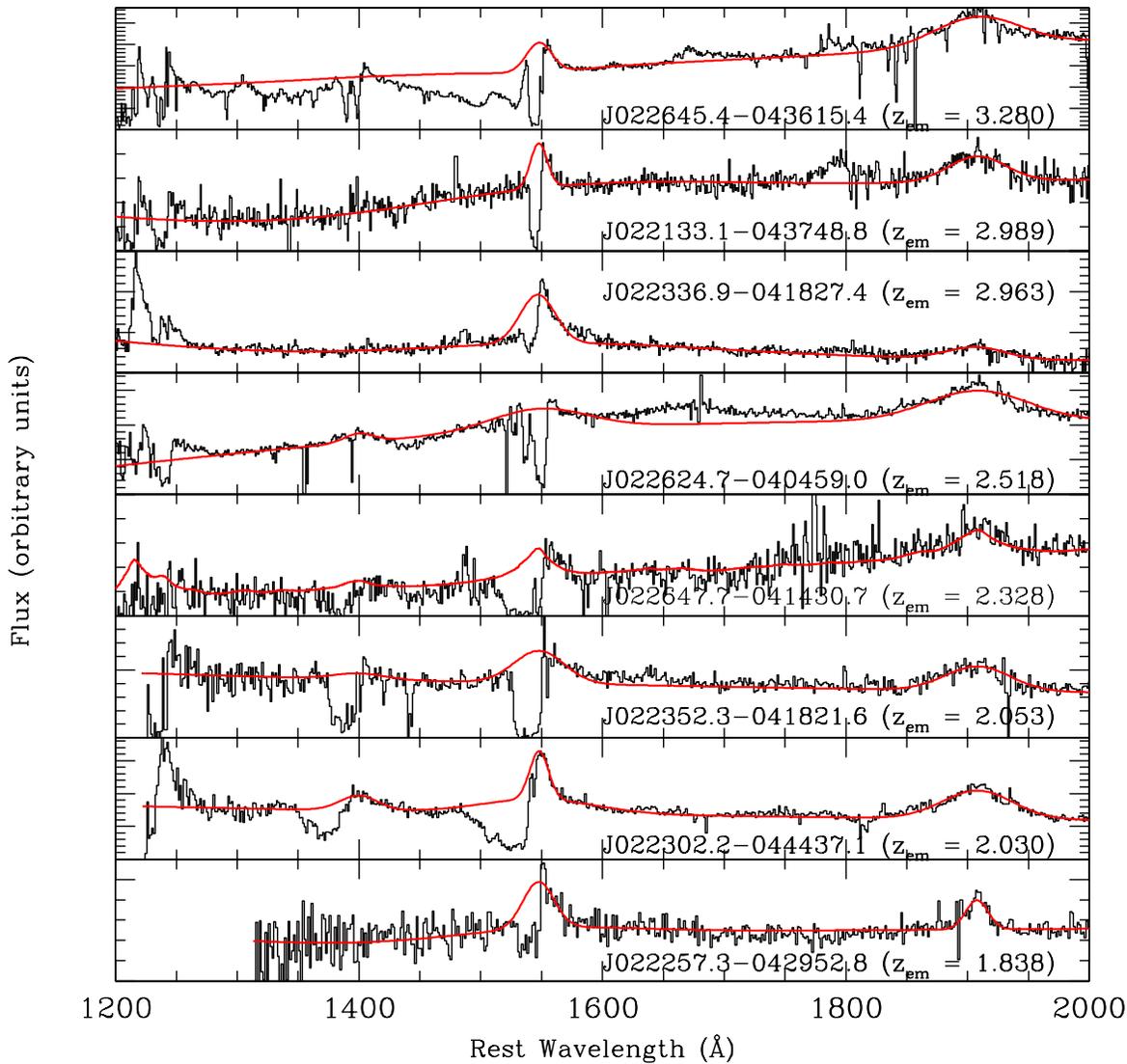,width=16cm}
\caption{The rest frame spectra of  BAL quasars detected by XMM in our 
sample. The best fitted continuum is over plotted.}
\label{xmmbal}
\end{figure*}

\begin{figure*}
\hspace*{-1.5cm}\psfig{file=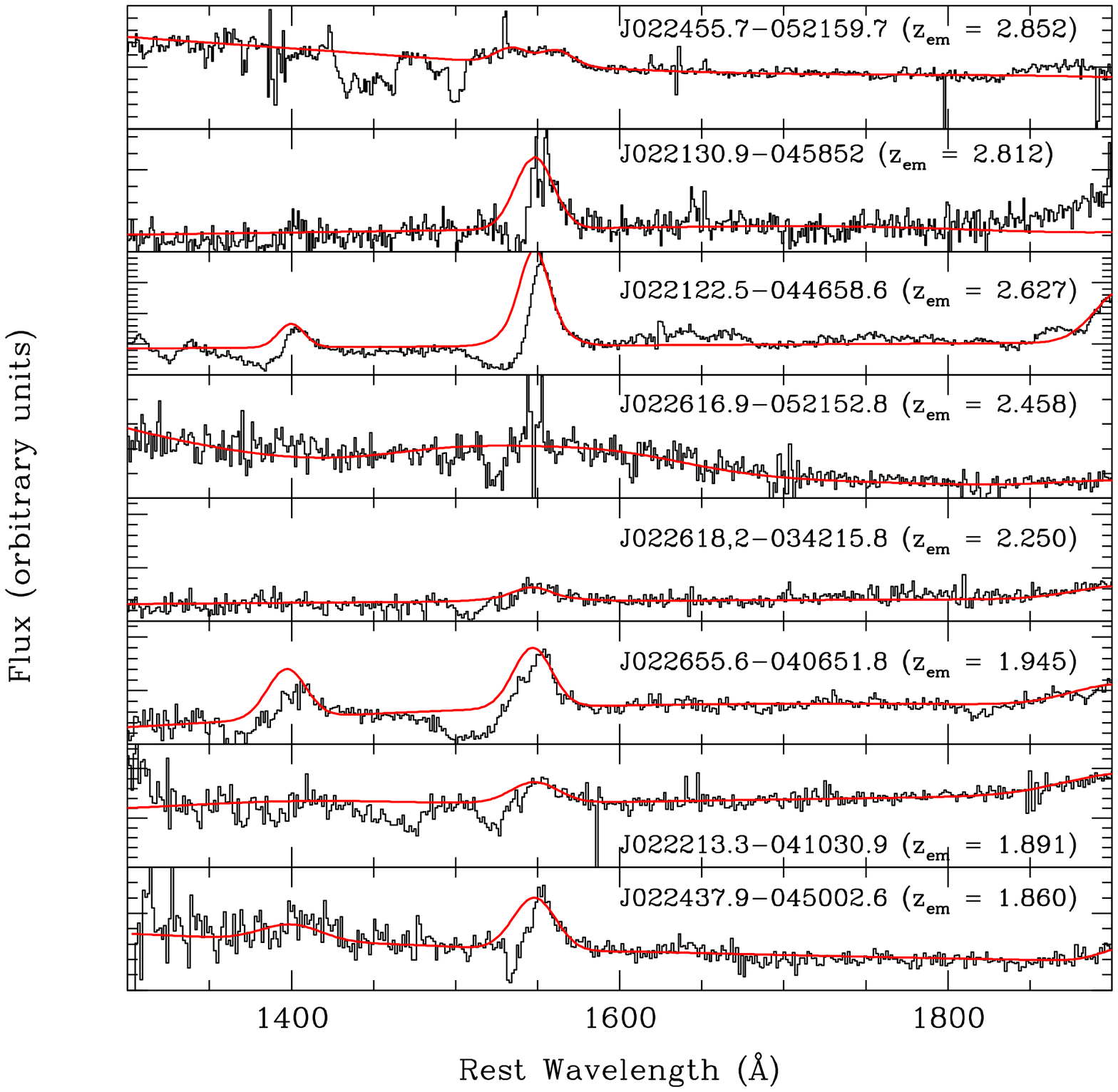,width=16cm}
\caption{The rest frame spectra of BAL quasars in our sample without 
X-ray detection. The best fitted continuum is over plotted. 
}
\label{xrayquietbal}
\end{figure*}

\section{Identifying BALs}
From our sample of 159 quasars, we initially identified 
quasars with broad C~{\sc iv} absorption close to the emission redshift by eye. 
We then for these quasars calculated the balnicity index (BI) as defined by 
Weymann et al. (1991) 
%{\bf as this is the most widely used method 
%to separate BAL quasars from non-BAL quasars}:
\begin{equation}
BI = -\int_{25000}^{3000}\left[1 - \frac{f(v)}{0.9}\right] Cdv.
\end{equation}
Here, $f (v)$ is the continuum-normalized flux at a velocity $v$ (in km~s$^{-1}$) 
defined with respect to the quasar rest frame. The dimensionless value $C$ is initially 
set to zero. It is set to 1.0 whenever the quantity in brackets has been continuously 
positive over an interval of 2000~km~s$^{-1}$. 
%It is reset to zero whenever the quantity
%in brackets becomes negative.
Traditionally BAL quasars are defined as quasars with BI~$>$~0.
%
%We first identified quasars that show  absorption lines close to the C~{\sc iv} emission line.  
We  fitted a smooth continuum to these  
quasar spectra using the method similar to the one 
described in Trump et al. (2006). The normalized spectrum is used to measure BI.  
We also simultaneously estimated $V_{max}$, the maximum outflow 
velocity at which $f(v)$ is 0.9.
The properties of the 16 new BAL (all HiBALs) quasars with BI greater than 100~km~s$^{-1}$
are summarized in  Table 1. They are shown as crosses (BALs detected
in X-rays in XMM-LSS) and open circles (BALs un-detected in X-rays
in XMM-LSS) in Fig. 1.
Of these 16 BAL quasars, we have Mg~{\sc ii} coverage for 6 BAL quasars. 
None of these 6 BALs are found to be LoBALs. This is not surprising, as
we would expect to see 1.6 LoBALs if all the 16 BALs had Mg~{\sc ii} 
coverage in their spectra, as only ~10\% of optically selected BALs
are known to be LoBALs.

We find a BAL quasar fraction 
$f_{\rm BAL}=N({\rm BAL})/N({\rm Total})$$\approx$10$\pm$3\% in  our sample. If we 
restrict ourselves to the optical colour selected quasars we find 8$\pm$3\%  of them are BAL 
quasars. We also find 7$\pm$5\% of the XMM selected quasars to be BAL quasars. This is consistent
with the rate found for the optical colour selected candidates. 
There are 12 quasars which are selected via IR colour selection but are not part of our optical colour 
selection. Out of these 12 quasars, 4 are found to be BAL quasars. We thus find a BAL detection rate 
of 33$\pm$19\% among the IR only selected quasars.

%{\bf BAL quasars suffer from (a) reddening in their UV spectra and (b) large
%absorption in their BAL troughs. Because of these two effects, BAL quasars
%get missed out in magnitude limites samples compared to normal 
%quasars and thus limiting us to have a clear idea of the true
%fraction of BAL quasars.
%A correction
%to these effects is therefore necessary to get the true BAL quasar
%fraction in optically selected quasar samples.} 
In our sample of 
 optically selected quasars, the observed BAL quasar fraction is very similar 
to the fraction 10$-$15\% reported in the literature from  optically 
selected quasar samples 
(see Weymann et al. 1991, Hewett \& Foltz 2003; Reichard et al. 2003; 
Tolea, Krolik \& Tsvetanov 
2002; Trump et al. 2006, Knigge et al. 2008, Gibson et al. 2009). 
However, the BAL quasar fraction found in optically selected sample could not
represent the true BAL fraction. This is because optical surveys will miss 
BAL quasars as their optical broad band colours are affected by 
reddening in their UV spectra and large absorption in their
BAL troughs. This is clearly evident in the 
higher BAL quasar 
fraction of 23$\pm$3\% 
reported by Dai et al. (2008) in the 2MASS selected quasars. 
When Hewett \& Foltz (2003) 
considered
the effect of reddening in their BAL quasars sample, they found
a BAL quasar fraction of ~22\% similar to  
that of Dai et al. (2008). 
High BAL quasar fraction is also claimed in the samples selected
based on radio emission (see Shankar et al. 2008; Urrutia et al. 2009). Interestingly, this is 
also consistent 
with the higher BAL quasar fraction we find among the objects selected from IR colours only.
Note that Dai et al. (2008) argued that the selection based on optical colours has significant selection biases 
against BAL  quasars, and the BAL quasar fraction in NIR colour selected sample more accurately reflects
the true fraction of BAL quasars. Indeed, Allen et al. (2010) have
shown that, when the incompleteness related to the BAL identification 
in the SDSS spectrum and the colour selection method used to identify
QSOs were taken into account, the actual BAL fraction can be as high as 41$\pm$5 per cent instead
of 8.0$\pm$0.1 per cent found from the SDSS sample.

%Recently, from a large sample of quasars
%in SDSS DR6, Allen et al. (2010) report an observed C~{\sc iv} BAL quasar fraction
%of 8.0 $\pm$ 0.1 per cent. This is similar to what we find from our optical
%colour selected quasar sample. However, they derive an intrinsic 
%C~{\sc iv} BAL quasar fraction as high as $\sim$41 $\pm$ 5 per cent after
%correcting the observed fraction for various selection effects.}

Of the 16 BAL quasars in our sample, 8 are detected in XMM-LSS (50$\pm$22\%).
Even if we restrict the sample to the colour selected objects, we find equal numbers of BAL quasars 
with and without detectable X-ray emission (i.e 50\%). The spectra of the 8 BAL quasars that 
are detected by XMM in XMM-LSS are shown in Fig.~\ref{xmmbal}. Fig.~\ref{xrayquietbal} shows
the spectra of the remaining 8 BAL quasars that are not detected in XMM-LSS.
Our X-ray detection rate is slightly lower than the 77$\pm$20\% reported by G06,  
74$\pm$13\% reported by Gibson et al. (2009) and  63$\pm$16\% reported by Fan et al. (2009). 
However, within uncertainties our value matches well with that from previous studies. There are
two main differences between our study and the above listed studies. 
Firstly, our sample is optically selected and then correlated with XMM-LSS, thereby, BAL and non-BAL
quasars have similar X-ray flux limits and secondly, 
the quasars in our study are typically fainter than that used in the other three studies (see Fig.~5). It is
also interesting to note that on an average, objects in the sample of G06 are optically
more luminous than those of Fan et al. (2009). Similarly, objects in our sample are systematically fainter
than the sample considered by Fan et al (2009). 

There are seven radio-loud quasars in our sample (see Table~5 of Stalin et al. 2010). None of them is 
a BAL quasar. Using the BAL fraction found by Shankar et al (2008) for radio selected quasars,
we should expect 1.5 radio-loud BAL quasars.

\section{Properties of BAL quasars in our sample}
\subsection{Optical-to-X-ray spectral index ($\alpha_{\rm ox}$)}

The broad band spectral index $\alpha_{\rm ox}$ is generally used to
quantify the relative UV to X-ray power of the quasar. 
%{\bf and this is the way to quantify the relative UV and X-ray power at 
%those two frequencies}.
%by assuming 
%that the rest frame flux emitted at 2500~\AA~ can be connected to 
%the one at 2~keV with a simple power law. 
We estimated $\alpha_{\rm ox}$ 
for each of the spectroscopically identified AGN 
%and ELGs (as some of the ELGs are powered by AGN) 
in our sample. 
For this we have converted the observed $i^{\prime}$-band 
magnitudes to fluxes following the definition
of the AB system (Oke \& Gunn 1983)

\begin{equation}
S_{\rm i^{\prime}} = 10^{-0.4 (m_{\rm i^{\prime}} + 48.60)}.
\end{equation}
where $S_{\rm i^{\prime}}$ and $m_{\rm i^{\prime}}$ are respectively the flux and 
magnitude in the i$^{\prime}-$band. 
%{\bf We note here, though it is ideal 
%to use the magnitues of the quasars measured in the filter close to the 
%restframe 2500 \AA, we decided to use the $i^{\prime}$-band as all quasars
%have magnitudes in available in this filter}.
The luminosity at the frequency corresponding to 2500~\AA~ in the rest-frame is calculated 
following Stern et al. (2000)
\begin{equation}
L_{\nu_1} = \frac{4 \pi D_{\rm l}^2}{(1+z)^{1+\alpha_{\rm o}}} (\frac{\nu_1}{\nu_2})^{\alpha_o} S_{\nu_2}
\end{equation}
where $\nu_2$ is the observed frequency corresponding to 
the $i^{\prime}-$band,  $S_{\nu_2}$ is the observed flux 
in $i^{\prime}-$ band, $\nu_1$ is the rest-frame frequency corresponding 
to 2500 \AA ~and $D_l$ is the luminosity distance. An optical spectral 
index $\alpha_{\rm o}$ = $-$0.5 (Anderson et al. 2007) is 
assumed ($S_{\nu}$~$\propto$~$\nu^{\alpha}$). 
The luminosity at 2~KeV in the rest frame is obtained using a similar 
equation as Eq.~3, assuming  a X-ray spectral index 
%({\bf however a 
%bit steeper}) 
of 
$\alpha_{\rm x}$~=~$-$1.5 (Anderson et al. 2007).  

Thus, the broad band spectral index $\alpha_{\rm ox}$ is obtained as
\begin{equation}
\alpha_{\rm ox} = \frac{{\rm log}(L_{2KeV}/L_{2500\AA})}{{\rm log}(\nu_X/\nu_{opt})}
\end{equation}
Here $L_{2 KeV}$ and $L_{2500 \AA}$ are the rest 
frame monochromatic luminosities  (in erg~s$^{-1}$~Hz$^{-1}$)
at $\nu_{\rm X} = 2 KeV$ and $\nu_{\rm opt}$ corresponding to
2500~{\AA} respectively.

\begin{figure}
\psfig{bbllx=47bp,bblly=177bp,bburx=273bp,bbury=417bp,file=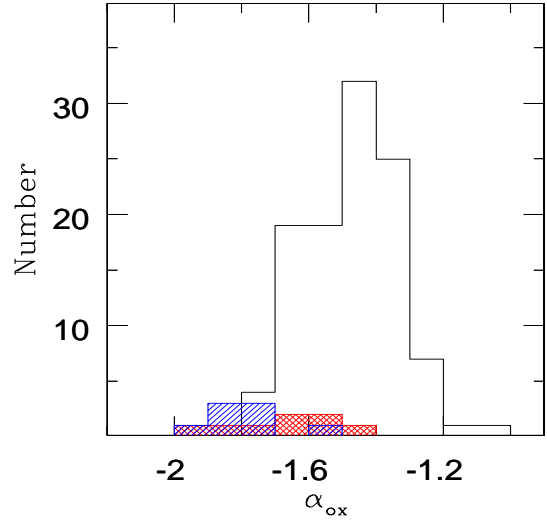,width=7.5cm,height=7.cm,clipp=}
\caption{Distributions of $\alpha_{\rm ox}$ in our sample. 
The solid histogram is  for the XMM detected non-BAL quasars. 
The hashed histogram  is for the XMM detected BAL quasars and the 
shaded histogram shows the upper limits of $\alpha_{\rm ox}$ for the 
X-ray un-detected BAL quasars.}
\label{alphaox_hist}
\end{figure}

Fig.~\ref{alphaox_hist} shows the distribution of observed $\alpha_{\rm ox}$ (not corrected for
intrinsic and Galactic absorption) for the sample of XMM detected non-BAL and BAL quasars. The BAL quasars 
have systematically lower values of $\alpha_{\rm ox}$ compared to the non-BAL quasars. 
We find a mean $\alpha_{\rm ox}$~=~$-$1.47$\pm$0.13 and $-$1.66$\pm$0.17 for the XMM detected 
non-BAL and BAL quasars respectively.  
%We also notice that 
%the upper limits on $\alpha_{\rm ox}$ obtained for the BAL QSOs
%with no X-ray detection in the XMM data are lower than the
%mean value of  $\alpha_{\rm ox}$ found for XMM detected non-BAL QSOs. 
%
In our sample,  the upper limits of $\alpha_{\rm ox}$  derived for the XMM 
un-detected BALs are towards the lower end of the distribution of $\alpha_{\rm ox}$
found for XMM detected non-BAL quasars as can be seen in Fig. 4.
%{\bf Though the XMM un-detected BALs have higher upper limits of $\alpha_{\rm ox}$ compared to XMM detected BALs, they are however, lower than the 
%mean value of $\alpha_{\rm ox}$ found for XMM detected non-BAL quasars}.
%The two parameter KS-test suggests that the probability for the BAL 
%quasars to be drawn from the same parent population as non-BAL quasars is only 0.9\%.

We now compare our results with other BAL quasar studies. 
In Fig.~5,  we show the distribution of $L_{2500\AA}$ luminosity and $\alpha_{\rm ox}$ for our sample of 
BAL quasars as well as other quasar samples published in literature. Objects in our sample 
are systematically fainter than those in the  published samples namely G06, Fan et al. (2009), Giustini 
et al. (2008)  and Gibson et al. (2009). The $\alpha_{\rm ox}$  distribution of our sample has some overlap with
Fan et al. (2009), Giustini et al. (2008) and Gibson et al. (2009) samples, but is offset from the distribution of 
G06. G06 found a median $\alpha_{\rm ox}$ of $-$2.20, which is 
much smaller than that found by Fan et al. (2009), Giustini et al. (2008) and Gibson et al. (2009). The objects 
in the G06 sample are luminous and the low value of $\alpha_{\rm ox}$  found might be due to the 
dependence of $\alpha_{\rm ox}$ on luminosity reported by
various authors (Green et al. 2009, Steffen et al. 2006, Just et al. 2007). 
Recently, for CFHTLS quasars, Stalin et al. (2010) found a  dependence of 
$\alpha_{\rm ox}$ on $L_{2500\AA}$  and is 
given by

\begin{figure}
\psfig{bbllx=47bp,bblly=177bp,bburx=555bp,bbury=680bp,file=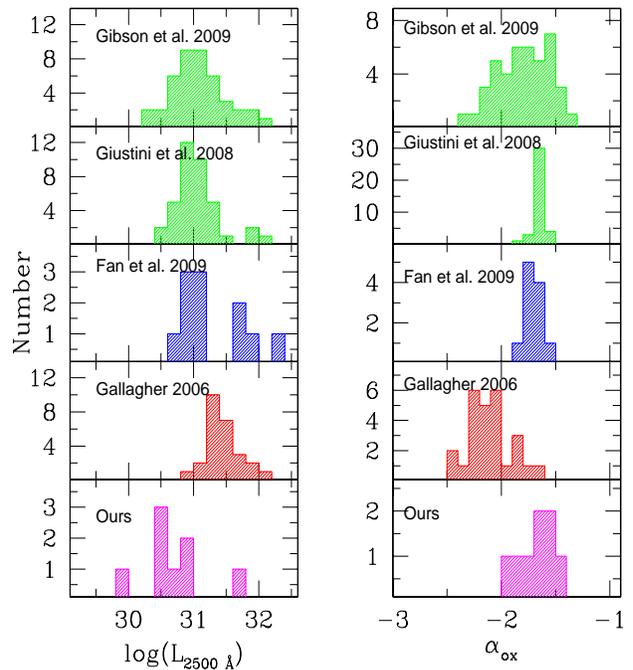,width=8.5cm,height=9cm,clipp=}
\caption{Distribution of the monochromatic luminosity at 2500 \AA ~(left panel) and $\alpha_{\rm ox}$ 
(right panel) in several BAL quasar samples. The references from where the samples are taken are given in each panels.}
\label{sampledist}
\end{figure}

\begin{figure}
\psfig{bbllx=53bp,bblly=181bp,bburx=279bp,bbury=500bp,width=8cm,height=8cm,file=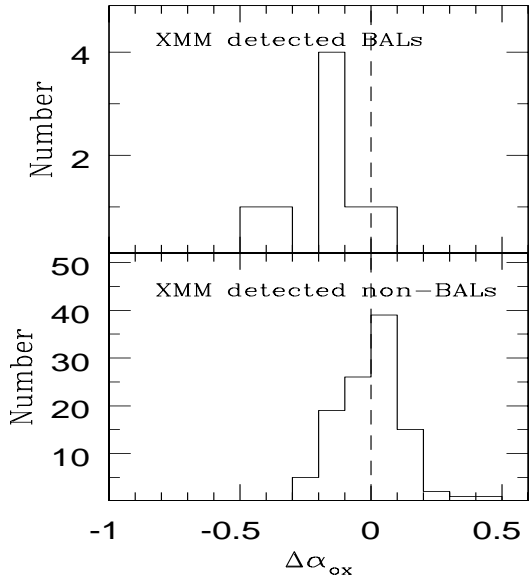,clipp=}
\caption{Distributions of $\Delta \alpha_{\rm ox}$ in our sample. The top panel 
is for XMM detected BAL quasars  and the bottom panel is for the XMM detected non-BAL quasars.
}
\label{aoxhist}
\end{figure}

\begin{equation}
\alpha_{\rm ox} = (-0.065\pm0.019) log(L_{2500}) +(0.509\pm0.560).
\label{eqalox}
\end{equation} 

This relation is in agreement to that found by Green et al. (2009), however, flatter than the value 
found by Steffen et al. (2006) and Just et al. (2007). To overcome this dependency 
of $\alpha_{\rm ox}$ on $L_{2500\AA}$ luminosity, we study the distribution of 
$\Delta \alpha_{\rm ox}$ values for our sample. For calculating $\Delta \alpha_{\rm ox}$ values, we opted
to use the relation between  $\alpha_{\rm ox}$ and $L_{2500\AA}$ obtained by Stalin et al. (2010) for the
CFHTLS quasars.
$\Delta \alpha_{\rm ox}$ is defined as the difference between the observed $\alpha_{\rm ox}$ and
$\alpha_{\rm ox}(L_{2500\AA})$ calculated using Eq.~\ref{eqalox} given the optical luminosity of the
object. This gives an estimate of how the X-ray luminosity of 
a BAL quasar differs from that of a typical non-BAL quasar of the same 
$L_{2500\AA}$. For BAL quasars culled from literature, $\Delta \alpha_{\rm ox}$ was
recalculated using Eq. 5.
The distributions of $\Delta \alpha_{\rm ox}$ for our sample of BAL and non-BAL quasars
are shown in Fig.~6. In the case of XMM detected non-BAL quasars, $\Delta \alpha_{\rm ox}$ is found 
to be distributed symmetrically around zero. 
On the contrary, the distribution of $\Delta \alpha_{\rm ox}$ for XMM detected BAL quasars is offset from zero.
The average $\Delta \alpha_{\rm ox}$ for XMM detected non-BAL and BAL quasars are 0.005$\pm$0.119 
and $-$0.169$\pm$0.161 respectively. 
Using Eq.~\ref{eqalox} we can say that $\Delta \alpha_{\rm ox}$ of $-0.169$ corresponds to 
an X-ray luminosity weaker by a factor of $\sim$3 with respect to typical non-BAL quasars of the same
$L_{2500\AA}$. Thus, in our sample, 
BAL quasars appear to be X-ray weak compared to their non-BAL counterparts. This is similar to what some 
previous studies have reported (G06; Fan et al. 2009; Gibson et al. 2009). 

\begin{table*}
 \centering
% \begin{minipage}{140mm}
\caption{Properties of BAL quasars in CFHTLS}
\begin{tabular}{cccrrrrccrr} \hline
RA         & Dec        & $g^{\prime}$    & $\alpha_{\rm ox}$~~~ & $\Delta\alpha_{\rm ox}$~~~ & $L_{\rm 0.5-2Kev}$ & $L_{\rm 2KeV}~~~$ & $L_{2500\AA}$ & $z$      & $V_{\rm max}$ & Balnicity\\ 
h:m:s      &  d:m:s     &  (mag)          &               &     &    (erg/s)~~~         &   (erg/sec/KeV)&  (erg/sec/\AA)&          & (km/sec)  &   index   \\ 
           &            &                 &                &    &                    &                &               &          &           &  \\ \hline
02:23:02.278 & $-$04:44:37.057  & 19.89 & $-$1.94  & $-$0.466  & 43.50          & 26.36      & 30.82          & 2.03 & 10335.0   & 3200.0    \\
02:23:52.537 & $-$04:18:21.620  & 20.54 & $-$1.83  & $-$0.354 & 43.51          & 26.37      & 30.54          & 2.05 &  5701.0   & 2126.0    \\
02:26:24.710 & $-$04:04:59.075  & 19.84 & $-$1.63  & $-$0.119 & 44.48          & 26.77      & 31.00          & 2.52 & 19409.0   &  500.0    \\
02:26:47.693 & $-$04:14:30.763  & 20.78 & $-$1.63  & $-$0.152 & 44.06          & 26.34      & 30.59          & 2.33 &  8233.0   & 6065.0    \\
02:26:45.443 & $-$04:36:15.475  & 19.92 & $-$1.72  & $-$0.174 & 44.84          & 27.17      & 31.66          & 3.28 & 22440.0   & 6567.0    \\
02:22:57.309 & $-$04:29:52.856  & 21.68 & $-$1.56  & $-$0.119  & 43.66          & 25.90      & 29.96          & 1.84 &  4308.0   &  580.0    \\
02:21:33.091 & $-$04:37:48.852  & 21.45 & $-$1.54  & $-$0.047 & 44.45          & 26.76      & 30.77          & 2.99 &  3942.0   &  109.0    \\
02:23:36.896 & $-$04:18:27.431  & 21.24 & $-$1.41  & 0.056 & 44.46          & 26.77      & 30.46          & 2.96 &  5569.0   &  479.0    \\
02:24:55.759 & $-$05:21:59.772  & 19.93 & $< -$1.94 & $< -$0.420  & $<$43.86          & $<$26.77      & 31.23          & 2.85 & 23766.0   & 3376.0    \\
02:26:16.910 & $-$05:21:52.820  & 20.39 & $< -$1.79 & $< -$0.309  & $<$43.70          & $<$26.59      & 30.65          & 2.45 &  6139.0   & 1742.0    \\
02:22:13.311 & $-$04:10:30.093  & 20.24 & $< -$1.84 & $< -$0.372  & $<$43.43          & $<$26.28      & 30.48          & 1.90 & 20640.0   & 2546.0    \\
02:24:37.908 & $-$04:50:02.572  & 20.42 & $< -$1.85 & $< -$0.375  & $<$43.41          & $<$26.26      & 30.46          & 1.87 &  4380.5   &  422.0    \\
02:21:22.497 & $-$04:46:58.606  & 20.43 & $< -$1.84 & $< -$0.345  & $<$43.78          & $<$26.67      & 30.88          & 2.63 & 10964.0   & 3620.0    \\
02:26:18.182 & $-$03:42:15.797  & 20.68 & $< -$1.77 & $< -$0.299  & $<$43.61          & $<$26.48      & 30.50          & 2.25 & 10053.0   & 4640.0    \\
02:21:30.978 & $-$04:58:52.086  & 21.95 & $< -$1.56 & $< -$0.105  & $<$43.85          & $<$26.75      & 30.22          & 2.81 &  4781.0   & 2996.0    \\
02:23:01.742 & $-$05:01:59.748  & 20.96 & $< -$1.78 & $< -$0.313  & $<$43.46          & $<$26.31      & 30.34          & 1.95 & 14430.0   & 6061.0    \\ \hline
\end{tabular}
%\end{minipage}
\label{baltable}
\end{table*}

\subsection{Correlation between X-ray and UV properties}
Models of BAL outflows based on radiation driven flows predict a correlation 
between the outflow velocity ($V_{\rm max}$)
and the quasar luminosity (Arav et al. 1994). The transfer of photon 
momentum depends on the column density and the ionization state of the 
absorbing gas. The ionization state of the BAL outflow very much 
depends on the optical to soft X-ray spectral energy distribution 
of the ionizing continuum (Srianand \& Petitjean 2000; 
Gupta et al. 2003). In order for an efficient 
transfer of radiative momentum to the absorbing gas it is important that
the gas is not over-ionized despite being very close to the central
engine (see Proga et al. 2000). This can be achieved if there is a 
highly ionized (not seen in UV absorption) gas component at the
base of the flow that shields X-rays from the UV absorbing gas
(Murray \& Chiang 1995). In such a scenario one would expect
the properties of the UV absorbing gas to be related to the 
properties of the X-ray absorbing gas at the base of the same flow.
We therefore investigate correlations between the kinematical properties of UV absorptions 
($V_{\rm max}$ and BI) with $\Delta\alpha_{\rm ox}$, 
an indicator of X-ray absorption, and $L_{2500\AA}$.
For these statistical study we add to
the new BAL quasars reported in this work, HiBALs from the 
literature. We have included 29 BAL QSOs from 
Gallagher et al. (2006), 38 BAL QSOs from Giustini et al. (2008),
42 BAL QSOs from Gibson et al.  (2009) 
and  35 BAL QSOs from Fan et al. (2009). Thus our extended sample 
consists of 160 high ionization BAL QSOs. Even though 
our new BAL quasars constitute only 10\% of the
extended sample, it is significant to the statistical analysis as
they are systematically fainter in optical luminosity
than the ones from the literature.

\begin{figure}
\psfig{bbllx=21bp,bblly=170bp,bburx=522bp,bbury=641bp,file=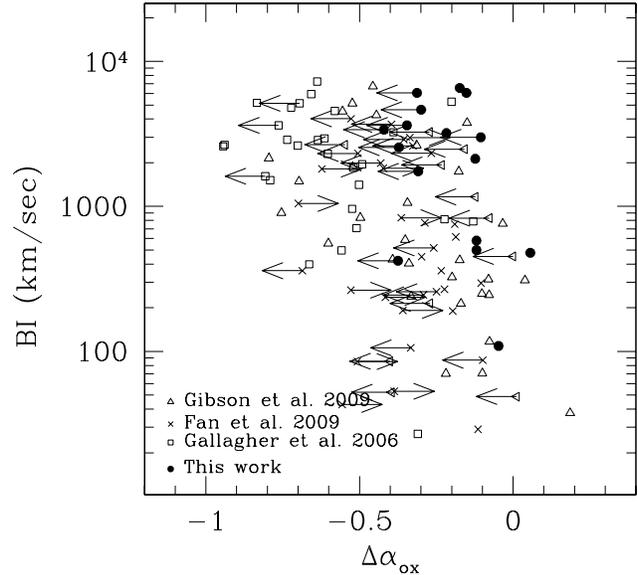,width=8.5cm,height=8cm,clipp=}
\caption{Correlation between BI and $\Delta \alpha_{\rm ox}$ for 
various BAL quasar samples taken from the literature.}
\label{biaox}
\end{figure}

In Fig.~\ref{biaox} we plot the balnicity index, BI, measured from the C~{\sc iv} absorption against 
$\Delta \alpha_{\rm ox}$ for the enlarged sample of BAL quasars including our own 16 BAL quasars.  
In our sample two out of three of
the BALs with BI greater than 5000~km/s show $-0.2\le \Delta\alpha_{\rm ox}\le-0.1$.
However, among objects with BI$\ge$2000~km/s, only 4 out of 10 BAL QSOs show
X-ray detection, whereas 4 out of 5 objects with 100$\le$BI$\le 1000$~km/s are
detected in X-rays. Thus there appears to be a tendency of QSOs with large
BI values to be weaker in X-rays. Now, we explore this correlation
in the extended sample discussed above. As BI values are not available for
the objects in Giustini et al. (2008) the corresponding objects are not considered in the 
analysis.  In Fig.~\ref{biaox} we plot $\Delta\alpha_{\rm ox}$ 
against BI for all the remaining 122 objects in the combined sample.
The Kendall - $\tau$ test as implemented in the Astronomy Survival Analysis (ASURV) 
package (Lavalley et al. 1992), that treats both upper and lower limits, gives a probability 
of $<$0.01\% that the two variables are uncorrelated (see Table.~\ref{baltable1}). 
This is consistent with a 99.95\% correlation found by Fan et al. (2009) between BI and 
$\Delta \alpha_{\rm ox}$ (see also Gibson et al. 2009). 
From Fig.~\ref{biaox}  it can be readily seen that most of the G06 sources have high BI occupying 
the top left  corner of the diagram. To check if the derived correlation is dominated by the G06 sources, 
we applied the test  without these sources. This time too,  a correlation is noticed, with the Kendall 
$\tau$ test giving a probability  of 0.01\% that BI and $\Delta \alpha_{\rm ox}$ are un-correlated (Table~\ref{baltable1}). This is mainly due to the fact that we have 
only upper limits on $\Delta \alpha_{\rm ox}$ in most QSOs with BI~$>$~1000~km/s. 

\begin{table}
 \centering
% \begin{minipage}{140mm}
\caption{Correlation analysis}
\begin{tabular}{llcr} \hline
Variables        & Npoints & \multicolumn{2}{c}{Kendall's $\tau$}  \\
                     &  & $\tau$   &  Prob.(\%)    \\ \hline
%$\alpha_{ox}$ v/s z         & 159 & 2.668     & 0.76    \\
$V_{\rm max}$ v/s $\Delta \alpha_{ox}$ & 160&5.243 & $<$0.01    \\  
$V_{\rm max}$ v/s $L_{2500}$ \AA & 160 & 4.421     & $<$0.01    \\
BI v/s $\Delta \alpha_{\rm ox}$  & 122 & 5.285     & $<$0.01    \\
BI v/s L$_{2500}$ \AA        & 122 & 2.405     &    1.62    \\
$V_{\rm max}$ v/s $\Delta \alpha_{\rm ox}$ & 131$^1$&5.274 & $<$0.01    \\ 
$V_{\rm max}$ v/s $L_{2500}$ \AA & 131$^1$ & 3.954     &    0.01    \\
$V_{\rm max}$ v/s $\Delta \alpha_{ox}$ & ~93$^2$&4.063 & $<$0.01    \\  
$V_{\rm max}$ v/s $L_{2500}$ \AA & ~93$^2$ & 3.694    & $<$0.01    \\
BI v/s $\Delta \alpha_{\rm ox}$  &  ~93$^1$ & 3.893     &    0.01    \\
BI v/s L$_{2500}$ \AA        &  ~93$^1$ & 0.624     &    0.53    \\
\hline
\end{tabular}
\begin{flushleft}
{$^1$ sample excluding data from G06.}

{$^2$ sample excluding data from G06 and Giustini et al. (2008).}
\end{flushleft}
%\end{minipage}
\label{baltable1}
\end{table}

\begin{figure}
\hspace*{-0.2cm}\psfig{bbllx=21bp,bblly=170bp,bburx=522bp,bbury=641bp,file=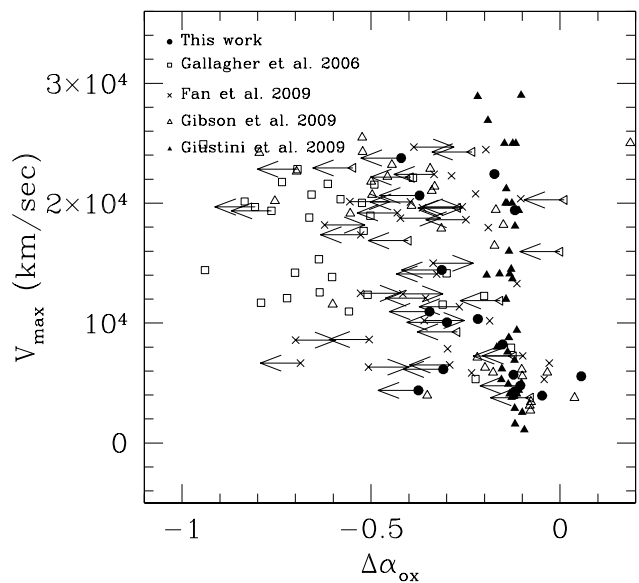,width=8.5cm,height=8cm,clipp=}
\caption{Maximum outflow velocity, $V_{\rm max}$ versus 
$\Delta \alpha_{\rm ox}$ for BAL quasar samples collected from the literature}. 
\label{vmax}
\end{figure}

Next we look for the correlations between  $V_{\rm max}$ and $\Delta \alpha_{\rm ox}$. There is no 
apparent trend between the two if we consider only our
sample. The objects with upper limits on $\Delta \alpha_{\rm ox}$ are spread
over the whole $V_{\rm max}$ range. In Fig.~\ref{vmax} we plot
$V_{\rm max}$ measured from the C~{\sc iv} absorption 
versus $\Delta \alpha_{\rm ox}$ in the extended 
BAL quasar sample. 
The Kendall $\tau$-test gives a probability of $<$ 0.01\% that the two variables are not correlated (Table~\ref{baltable1}). 
In the past, different samples have given different
results for this correlation. Gallagher et al (2006) and Gibson et al. (2009)
have found significant correlation between $V_{\rm max}$  and  $\Delta \alpha_{\rm ox}$ 
but this was not confirmed by Fan et al. (2009) and Giustini et al.
(2008). In the extended sample, even when we remove the points from Gallagher
et al. (2006), we do find a very small Kendall probability  (P $<$ 0.01\%)
that the two parameters are not correlated (see Table~\ref{baltable1}). 
If  we
remove points from Giustini et al. (2008), because they are not
selected homogeneously, 
the kendall $\tau$ test gives a probability
$<$ 0.01\% that $V_{max}$ and $\Delta\alpha_{ox}$ are uncorrelated.

It is a fact that the points in  Figs.~\ref{biaox} and \ref{vmax} appear to be scattered even though
the Kendall test hints to the existence of a possible correlation between 
$\Delta\alpha_{\rm ox}$ and kinematical parameters derived from the UV absorptions. This is mainly
due to the paucity of points in the bottom left corner of both the plots.
\begin{figure}
\hspace*{-0.2cm}\psfig{bbllx=21bp,bblly=170bp,bburx=522bp,bbury=641bp,file=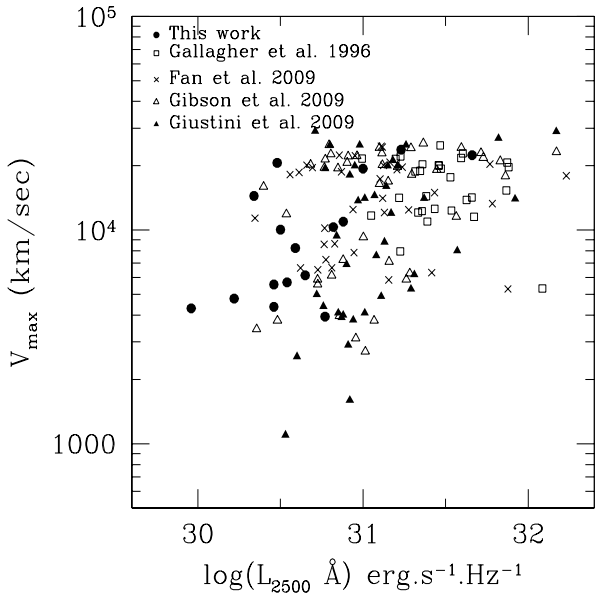,width=8cm,height=8cm}
\caption{Maximum outflow velocity, $V_{\rm max}$, versus the monochromatic luminosity at 
2500~\AA ~for various BAL quasar samples collected from the literature.}
\label{lvmax}
\end{figure}
To explore this further, we plot in Fig.~\ref{lvmax},  $V_{\rm max}$ measured from the C~{\sc iv} 
absorption against $L_{2500\AA}$. 
The kendall $\tau$-test gives a probability  $<$0.01\% that $V_{\rm max}$ and $L_{2500\AA}$ are not 
correlated both for the whole sample, sample excluding G06 sources, and
the sample excluding both G06 and Giustini et al. (2008) sources.
BAL quasars with high UV luminosities tend to have higher values of $V_{\rm max}$. 
While there is a large range in $V_{\rm max}$, for a given  $L_{2500\AA}$ at
log($L_{2500\AA}$)~$<$~31, there is a lack of objects
with $V_{\rm max}$~$<$~10$^4$~km/s at the high luminosity end. Thus, there
seems to be a lower envelope, the presence of which dominates the above correlation.
This seems to be also the case when we plot BI against $L_{2500\AA}$ (see Fig.~\ref{bivl}).
The marginal correlation found between the two parameters may mainly be dominated
by the absence of low BI sources at high optical luminosity. 
The above results are also consistent with that found by
Ganguly et al. (2007) 

A relationship between $V_{\rm max}$ and luminosity is expected if the broad troughs originate 
in a radiatively driven wind. From the analysis of C~{\sc iv} absorptions in low-$z$ Seyfert galaxies
and high-$z$ high luminosity QSOs,  Laor \& Brandt (2002) have shown that radiation-pressure
force multiplier increases with luminosity  (see also Ganguly et al. 2007). 
If true, this could explain the lower-envelopes seen here. However, our analysis also shows that 
there is a large scatter in BI and $V_{\rm max}$ at  any given luminosity. 
This probably means that 
even if  radiation pressure  is the main driver of the flow,  
other parameters are important such as the launching radius, the shape of the ionizing
spectrum, the mass of the wind, the properties of the confining medium, etc.. 

\begin{figure}
\hspace*{-0.2cm}\psfig{bbllx=21bp,bblly=170bp,bburx=522bp,bbury=641bp,file=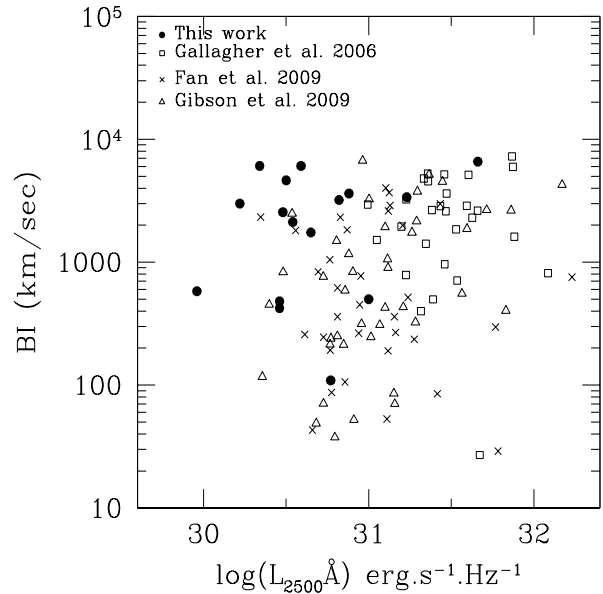,width=8cm,height=8cm}
\caption{The balnicity index, BI, is plotted versus the QSO monochromatic luminosity
at 2500~\AA ~for different samples of BAL quasars collected from the literature.}
\label{bivl}
\end{figure}

\subsection{X-ray hardness ratio}
\begin{figure}
\psfig{bbllx=55bp,bblly=190bp,bburx=290bp,bbury=411bp,file=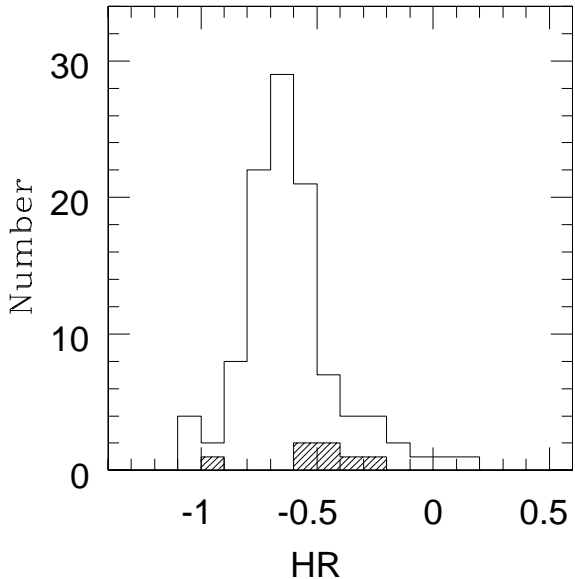,clipp=}
\caption{Distributions of the X-ray hardness ratio of the XMM-detected
non-BALs (solid histogram) and BAL quasars (shaded histogram)}
\label{hrfig}
\end{figure}

The hardness ratio is often the only spectral information that is possible to extract when the X-ray source is weak.
We queried the observations available in the XMM-Newton Science Archive version 6.0 (XSA) for 
all objects in our sample. For each observational ID, XSA provides the identified sources
along with count rates and their associated errors in five energy bands. Thus, using XSA, we extracted 
the count rates for all our XMM-detected non-BAL and BAL quasars and estimated the hardness ratio (HR)
between the soft (0.2$-$2 keV) and hard (2.0$-$10.0 keV) bands as:

\begin{equation}
HR = \frac{C(2.0-12.0 {\rm keV}) - C(0.2 - 2.0 {\rm keV})}{C(2.0-12.0 {\rm keV}) + C(0.2 - 2.0 {\rm keV})}
\end{equation}

where C(2.0$-$12.0 keV) and C(0.2$-$2.0 keV) are, respectively, the count rates in the 2.0$-$12.0 keV 
and 0.2$-$2.0 keV bands. The errors in HR were estimated by propagating the errors in each individual 
energy bands. In our sample we have a total of 109 XMM detected non-BAL  quasars. Of these the 
count rates were available for 106 sources. The average HR for these 106 sources is 
$-$0.61$\pm$0.20. Similarly, of the 8 XMM detected BAL quasars, good count rates are available for 
7 XMM BAL quasars. The average HR for these 7 BAL quasars is $-$0.48$\pm$0.20. The HR distributions 
for the XMM detected BAL and non-BAL quasars are shown in Fig.~\ref{hrfig}. As the number of BAL 
quasars with X-ray detection is small, the trend of high HR in BAL quasars cannot be confirmed with any
statistical significance.

\subsection{X-ray spectral analysis}
\begin{figure*}
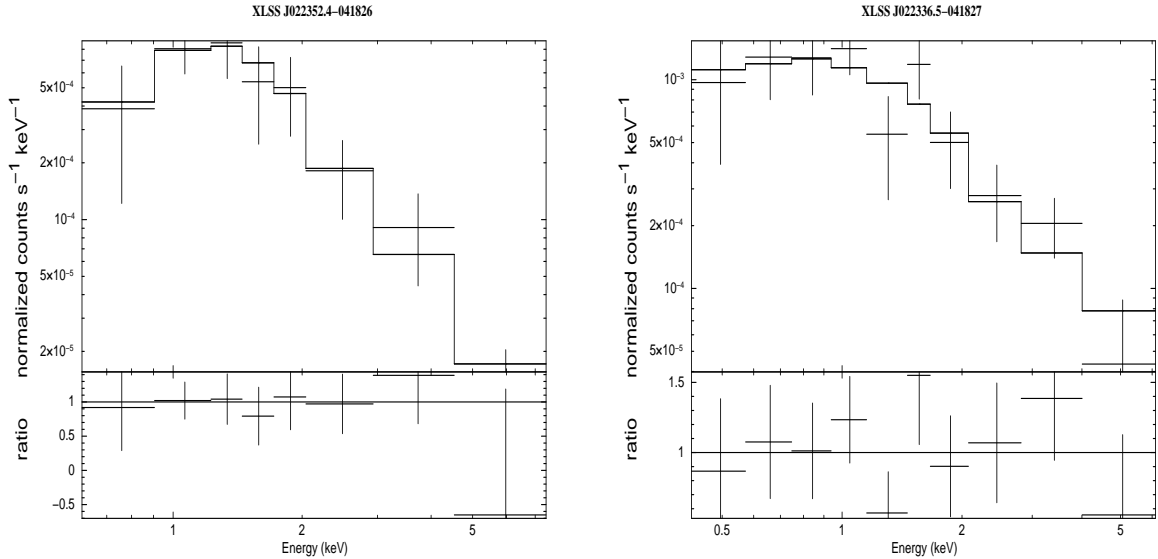

\hbox{
\psfig{file=id02.eps,width=8cm,height=8cm,angle=-90}
\psfig{file=id08.eps,width=8cm,height=8cm,angle=-90}
}
\caption{ The observed and the fitted spectra of the two XMM detected BAL quasars with 
counts larger than 100.
}
\label{xspec}
\end{figure*}

We also tried to perform a spectral analysis of archival XMM observations available for the
8 BAL quasars. For this we retrieved from XSA the original observational data file (ODF) for the 8 sources. 
These data were then reduced using the XMM Science Analysis Software (SAS) verion 8.0.1.
The original events files were filtered to retain only single and double pixel events 
(PATTERN~$\le$~4) as well as to retain only good quality  events (FLAG=0).
From this filtered events file, time stamps having high background events
were also removed. From the final cleaned events file, individual source
spectra were generated for each XMM detected BAL quasars with an extraction 
radius of 15$-$20$^{\prime\prime}$ for the source and the background region.
The background region was extracted from noise free regions as close to
the source as possible. Also, ancillary response file (ARF) and redistribution
matrix file (RMF) at the position of the source were generated with the
{\it argen} and {\it rmfgen} tasks. Total counts for each of these 8 sources were
extracted in the 0.2$-$8.0 keV region. Of these 8 sources, only 2 sources have
counts greater than 100, suitable for a reasonable spectral analysis.
Therefore, for these 2 sources, spectra were grouped in 15 counts/bin
and taken into XSPEC (Arnaud 1996) for spectral analysis.  
We performed the fit by minimizing $\chi^2$.
We model the spectra 
as a single power law continuum emission with absorption. The observed and
fitted spectra are shown in Fig.~\ref{xspec}. For both sources,  
$N_{\rm H}$ values are found to be less than $10^{22}$~cm$^{-2}$ and similar to the
Galactic values. Similar conclusion was found by 
Giustini et al. (2008) when they fitted the spectra of their BAL sample
assuming neutral absorbers. Recently, Streblyanska et al. (2010) found that
about 36\% of their sample of BALs have low $N_{\rm H}$ when the spectra
are fitted with a model including a neutral absorber. However, when 
the spectra are modeled with an ionized absorber, Streblyanska et al. (2010) 
found high values of $N_{\rm H}^i$  for more than 90\% of 
their sources. 
Due to the low count rates of the objects we study here, we are 
unable to apply  such detailed models. Therefore, the $N_{\rm H}$
values reported here for the two X-ray brightest objects in our 
sample should be treated as lower limits.
For our two sources, we found a photon index $\Gamma$ values, 1.62$\pm$0.44 
and 2.80$\pm$0.89, which is 
similar to that of typical radio-quiet quasars (Piconcelli et al. 2005) and 
within error of the mean photon index, $\Gamma$~=~1.87$\pm$0.21, found from
spectral analysis of 22 BAL quasars by Giustini et al. (2008).
The results of our X-ray analysis on the 8 BAL quasars are summarized in Table~\ref{table3}.

\begin{table*}
 \centering
% \begin{minipage}{140mm}
\caption{X-ray spectral properties of the XMM detected BAL quasars. Column 1 is the XMM name, column 2 is the XMM observation ID, 
column 3 is the hardness ratio measured between 0.2$-$2.0 keV and 2.0$-$12.0 keV, column 4 is the Galactic neutral column density 
%calculated from the Chandra web tool, 
derived from Dickey \& Lockman (1990), column 5 is the neutral column density from the spectral fit and last column gives 
the fitted photon spectral index}
\begin{tabular}{cccccc} \hline
Name                 &  ObsID      & HR       &  $N_{\rm H, Gal}$ ($10^{20}$)  & $N_{\rm H}$ ($10^{20}$) & $\Gamma$  \\ 
                     &             &           &  cm$^{-2}$                 &  cm$^{-2}$        &        \\ \hline
XLSS J022302.3-044437 & 0109520501 &    0.63 $\pm$ 1.09 & 2.68 &                 &                  \\
XLSS J022352.4-041826 & 0210490101 & $-$0.52 $\pm$ 0.17 & 2.54 &  6.04 $\pm$ 0.33 & 2.80 $\pm$ 0.89\\
XLSS J022624.7-040458 & 0112680201 & $-$0.28 $\pm$ 0.33 & 2.65 &                 &                 \\
XLSS J022647.7-041426 & 0112680101 & $-$0.33 $\pm$ 0.27 & 2.66 &                 &                 \\
XLSS J022645.4-043615 & 0112681301 & $-$0.41 $\pm$ 0.23 & 2.46 &                 &                 \\
XLSS J022257.2-042953 & 0109520601 & $-$0.50 $\pm$ 0.55 & 2.63 &                 &                 \\
XLSS J022132.9-043750 & 0112680801 & $-$0.90 $\pm$ 0.59 & 2.58 &                 & \\
XLSS J022336.5-041827 & 0210490101 & $-$0.45 $\pm$ 0.10 & 2.55 &  0.97 $\pm$ 0.11 & 1.62 $\pm$ 0.44                          \\ \hline

\end{tabular}
%\end{minipage}
\label{table3}
\end{table*}

\section{Conclusion}
We have presented a new sample of 16 BAL quasars selected from
a homogeneous sample of 159 $z_{\rm em} > 1.5$, $g^{\prime} < 22$ mag quasars
found in a region of CFHTLS overlapping with the XMM-LSS and SWIRE surveys.
\begin{enumerate}
\item We find a BAL quasar fraction of $\sim$10\% in the whole sample and of
$\sim$8\% among optically selected quasars (120 quasars). 
This is similar to what is found from other optically selected quasar samples (Weymann et al. 1991; 
Hewett \& Foltz 2003; Trump et al. 2006; Knigge et al. 2008; Gibson 
et al. 2009). Also, 7\% of the XMM selected quasars are found to be BAL quasars. If we 
consider the 12 quasars which were selected from SWIRE colours only (and rejected by the optical colour
selection), 4 are found to be BAL quasars which gives a BAL fraction of $\sim$33\%.  This agrees 
with the large 23\% BAL quasar fraction found in  2MASS selected quasars by 
Dai et al. (2008). We note here that recently Allen et al. (2010), after
correcting for selection biases, report
an intrinsic C~{\sc iv} BAL quasar fraction of $\sim$41$\pm$5 per cent, 
though their observed fraction is 8.0$\pm$0.1 per cent.
\item The values of $\Delta \alpha_{\rm ox}$, the deviation of the spectral index, $\alpha_{\rm ox}$, 
from the mean in the overall  sample at the same UV luminosity,
are distributed symmetrically around zero, whereas the $\Delta \alpha_{\rm ox}$
values of XMM detected BAL quasars are shifted towards lower values with an 
average of $-$0.169$\pm$0.161. This shows that the X-ray detected BAL quasars
are weaker than the X-ray detected non-BAL quasars by a factor of
about 3. This X-ray weakness of BAL quasars compared to non-BAL quasars
is similar to what was found in earlier studies (G06; Fan et al. 2009; 
Gibson et al. 2009) however, contrasts with the results of Giustini et al. (2008). 
The discrepancy we find is however
much less than what was reported by G06 who found that optically bright BAL quasars, 
are X-ray sources weaker by a factor of 30 compared to non-BAL quasars.
This might be due to the fact that G06 had  
used {\it Chandra} observations reaching significantly fainter 
flux levels than the XMM-Newton observations used here for 
our sample of BALs.
\item 
We investigated various correlations between the properties of 
the C~{\sc iv} absorptions and $\alpha_{ox}$ using an extended sample
gathered after combining our data with 
those of Gallagher et al. (2006), Giustini et al. (2008), Gibson et al. (2009) and Fan et al. (2009). 
For this large HiBAL quasar sample, $\Delta \alpha_{\rm ox}$ was calculated and
we find it to be correlated with the balnicity index, BI, and 
the maximum velocity of the outflow, $V_{\rm max}$. Similarly,
$V_{\rm max}$ and BI are correlated with the 2500~\AA~ monochromatic luminosity, $L_{2500\AA}$.
This suggests that quasars with high velocity outflows are X-ray weak. 
While there is a large range in $V_{\rm max}$, for a 
given  $L_{2500\AA}$ at log($L_{2500\AA}$)~$<$~31, there is a lack of objects
with $V_{\rm max}$~$<$~10$^4$~km/s at the high luminosity end. Thus there seems
to be a lower envelope, the presence of which dominates the above correlation.
This probably means that even if  radiation pressure  is the main driver of 
the flow, other parameters are important such as the launching radius, the 
shape of the ionizing spectrum, the mass of the wind, the properties of the 
confining medium, etc. 
\item We find the mean X-ray hardness ratio, HR, of XMM detected non-BAL 
and BAL quasars to be $-$0.61$\pm$0.20 and $-$0.49$\pm$0.20 respectively.
While this is consistent with the assumption that the X-ray weakness is due to
soft X-ray absorption, the number of X-ray detected BAL quasars is too small to make 
any statistically significant claim.

\item The fit of the X-ray spectra from two BAL quasars
shows that they have neutral hydrogen column densities smaller than $10^{22}$~cm$^{-2}$ and close to the 
Galactic values. This is based on a fully covering neutral absorber
model. However, if the absorber is ionized and/or partially covering the
X-ray source, the column density derived for these two BAL quasars 
should be considered lower limits. 
For these two BAL quasars, the photon spectral index is 
found to be similar to that of radio-quiet quasars. This agrees 
with the photon index values found recently from X-ray spectral analysis of  
a larger sample of BAL quasars (Giustini et al. 2008; Streblyanska 
et al. 2010). 

\end{enumerate}

\section*{Acknowledgments}
We thank the anonymous referees for their valuable comments which helped
to significantly improve the paper.
We also thank all the present and former staff of the Anglo-Australian Observatory for their work in building
and operating the AAOmega facility. This work used the CFHTLS data products, which is based on observations
obtained with MegaPrime/MegaCam, a joint project of CFHT and CEA/DAPNIA, at the 
Canada-France-Hawaii Telescope (CFHT) which is operated by the National Research Council (NRC) of Canada,
the Institut National des Science de l'Univers of the Centre National de la Recherche Scientifique (CNRS)
of France, and the University of Hawaii. This work is based in part on data products produced at TERAPIX
and the Canadian Astronomy Data Center as part of the Canada-France-Hawaii Telescope Legacy Survey, a collaborative
project of NRC and CNRS.

\end{document}